\documentclass[aps,prd,a4paper,twocolumn,showpacs,nofootinbib]{revtex4}
\usepackage{amssymb,amsmath,latexsym,mathrsfs}
\usepackage{graphicx,subfigure}
\usepackage{epsfig}
\usepackage{varioref,xr-hyper}
\usepackage{color}

\usepackage[hypertex,draft=false,verbose=false,debug=false
            hyperindex=true,hypertexnames=true]{hyperref}
\begin{document}

\newcommand{\nhat}{\hat{\bf n}}
\newcommand{\kvec}{{\bf k}}
\newcommand{\khat}{\hat{\bf k}}
\newcommand{\od}{\Omega_{\rm de}}
\newcommand{\omegae}{\Omega_{\rm e}}
\newcommand{\ome}{\Omega_{\rm e}}
\newcommand{\om}{\Omega_m}
\newcommand{\aeq}{a_{\rm {eq}}}

\newcommand{\Mpc}{{\rm ~Mpc}}
\newcommand{\sech}{{\rm ~sech~}}
\newcommand{\Tr}{{\rm ~Tr~}}

\title{Constraining Variations in the Fine Structure Constant in the presence of Early Dark Energy}

\author{Erminia Calabrese$^{a}$}
\author{Eloisa Menegoni$^{a}$}
\author{C.J.A.P. Martins$^{b,c}$}
\author{Alessandro Melchiorri$^{a}$}
\author{Graca Rocha$^{d}$}

\affiliation{$^a$ Physics Department and INFN, Universit\`a di Roma ``La Sapienza'', Ple Aldo Moro 2, 00185, Rome, Italy}
\affiliation{$^b$ Centro de Astrof\'{\i}sica, Universidade do Porto, Rua das Estrelas, 4150-762 Porto, Portugal}
\affiliation{$^c$ DAMTP, University of Cambridge, Wilberforce Road, Cambridge CB3 0WA, United Kingdom}
\affiliation{$^d$ Jet Propulsion Laboratory, California Institute of Technology, California, USA}

\begin{abstract}
We discuss present and future cosmological constraints on variations of the fine structure
constant $\alpha$ induced by an early dark energy component having the simplest allowed (linear) coupling to electromagnetism. We find that current cosmological data show no variation of the fine structure constant at recombination respect to the present-day value,
with $\alpha / \alpha_0 = 0.975 \pm 0.020$ at $95 \%$ c.l., constraining the energy density in early dark energy to $\Omega_e < 0.060$ at $95 \%$ c.l.. Moreover, we consider constraints on the parameter quantifying the strength of the coupling by the scalar field. We find that current cosmological constraints on the coupling are about 20 times weaker than those obtainable locally (which come from Equivalence Principle tests). However forthcoming or future missions, such as Planck Surveyor and CMBPol, can match and possibly even surpass the sensitivity of current local tests.
\end{abstract}

\maketitle

\section{Introduction}

Cosmology has recently entered a precision, data-driven era. The availability of ever larger, higher-quality datasets has led to the so-called concordance model. This is a remarkably simple model (with a small number of free parameters) which provides a very good fit to the existing data (see e.g. \cite{wmap7,act,Percival:2009xn,hst}). However, there is a price to pay for this success: the data suggests that $96 \%$ of the contents of the universe are in a still unknown form, that so far has never been seen in a laboratory. This is often called the dark sector of the universe.

Current best estimates suggest that this dark sector is in fact a combination of two distinct components (see e.g. \cite{dereview} for recent reviews). The first is called dark matter (making up about $23 \%$ of the universe) and it is clustered in large-scale structures like galaxies. The second, which has gravitational properties very similar to those of the cosmological constant first proposed by Einstein, is called dark energy and currently dominates the universe, with about $73 \%$ of its density. Understanding what constitutes this dark energy is one of the most important problems of modern cosmology. In particular, we would like to find out if it is indeed a cosmological constant, since there are many possible alternatives. These alternative models often involve scalar fields, an example of which is the Higgs field which the LHC is searching for.

There are two main differences between the cosmological constant and the models involving scalar fields (which are often collectively called dynamical dark energy models). First, in the former case the density of dark energy is a constant while in the latter the dark energy density does change. Second, the scalar fields will necessarily couple to other fields (unless they are prevented from doing so by symmetry principles) and thus can lead to further distinguishing features \cite{Carroll} such as variations of nature's fundamental couplings. A case in point is the coupling of the scalar field to the electromagnetic sector of the model, which will lead to spacetime variations of the fine-structure constant $\alpha$.

One way to distinguish a cosmological constant from dynamical dark energy is therefore to find several independent ways to measure the dark energy density (or its equation of state $w=p/\rho$) at several epochs in the universe, while at the same time searching for variations of $\alpha$ or other dimensionless couplings \cite{Couplings1}. 

Cosmic Microwave Background Anisotropies (CMB hereafter) have provided in the past years a powerful method to constrain variations in the fine structure constant in the early universe,
at the epoch of recombination (see e.g. \cite{avelino,Rocha,ichikawa,jap,petruta,menegoni,landau}) at the level
of $\sim 1 \%$. In the most recent analysis, parametrizing a variation in the fine structure constant as $\alpha/\alpha_0$, where
$\alpha_0=1/137.03599907$ is the standard (local) value and $\alpha$ is the value during the recombination process, the authors of \cite{menegoni} used the five year WMAP data, finding the constraint $0.987\pm0.012$ at $68 \%$ c.l..

The presence of a scalar field at recombination could induce variations in the fine structure constant. While the effects of a cosmological constant at recombination are completely negligible, dynamical scalar fields could track the dominant energy component, be present at recombination and induce variations in $\alpha$ if coupled to the electromagnetic sector. It is therefore timely to check for variations of the fine structure constant in cosmological data allowing 
at the same time the presence of an ``early" dark energy component at recombination. 
All recent cosmological constraints on $\alpha$ have indeed been obtained under the assumption of a cosmological constant or of a dark energy component with constant equation of state $w$
(\cite{menegoniw}), i.e. in the case of a negligible dark component at recombination. 
Clearly, if a dark energy is present in the early universe, degeneracies between
the parameters may arise and change the current results.
Moreover, in the past year, a significant experimental progress in determining the CMB angular power spectrum has been made, most notably from the Atacama Cosmology Telescope (ACT) (\cite{act}). This new data, as we will see in the next sections,  substantially improve
previous constraints on $\alpha$. 

In this paper we indeed follow this way of reasoning and we produce new constraints on variation of $\alpha$ allowing at the same time for the presence of dark energy  at recombination. In order to be as general as possible, we describe the scalar field adopting a phenomelogical approach, considering an early dark energy model (hereafter, EDE) as discussed in \cite{Doran:2006kp}. We consider a class of models where the link between the dark energy and electromagnetic sectors is explicit. Specifically we discuss present and future constraints on EDE models having the simplest allowed (linear) coupling to electromagnetism.

Our paper is therefore structured as follows:
in the next Section we briefly describe the early dark energy model considered and its coupling to electromagnetic sector. In Section III we present the data analysis method used while in Section IV we discuss our results.

\section{Early Dark Energy and the Electromagnetic Sector}

As discussed in the previous section we aim to place constraints on variations
of the fine structure constant with a non-negligible scalar field at recombination.
We describe the scalar field with a EDE model \cite{Doran:2006kp} 
where the dark energy density parameter and equation of state are parametrized in the following way : 

\begin{eqnarray}
\od(a) &=&  \frac{\od^0 - \omegae \left(1- a^{-3 w_0}\right) }{\od^0 + \Omega_{m}^{0} a^{3w_0}} + \omegae \left (1- a^{-3 w_0}\right) \label{eq:edeom}\\
w(a) &=& -\frac{1}{3[1-\od(a)]} \frac{d\ln\od(a)}{d\ln a} + \frac{a_{eq}}{3(a + a_{eq})} 
\label{eq:edew} 
\end{eqnarray}
where $\aeq$ is the scale factor at matter-radiation equality and $\od^0$ and $\Omega_{m}^{0}$ are the current dark energy and matter density, respectively. A flat Universe is assumed so $\Omega_{m}^{0}+\od^0 = 1$ and the present value for the equation of state is obtained from demanding $w(a=1)=w_0$. The energy density $\od(a)$ has a scaling behaviour evolving with time and going to a finite constant $\ome$ in the past.

The dark energy equation of state $w(a)$ follows the behaviours of the dominant component at each cosmic time : $w\approx 1/3$ 
during radiation domination, $w\approx 0$ during matter domination, and $w \approx w_0$ in recent epochs as in a cosmological 
constant era.  We add dark energy perturbations as in \cite{Calabrese:2010uf} by considering the EDE clustering properties through the effective sound speed $c_s^2 = \delta p/ \delta \rho$ and a viscosity parameter $c_{vis}^2$ accounting for the presence of anisotropic stresses. In the present analysis we assume these clustering parameters as constant with $c_s^2=1$ and $c_{vis}^2=0$ as expected in the case of a scalar field. For simplicity we also fix $w_0=-1$ since this value is well constrained by low redshift data. 

In any realistic dynamical scalar field scenario, the scalar field should be coupled to
the rest of the model, unless one postulates a (yet unknown) symmetry to suppress the
coupling. We are presently interested in the coupling between the scalar field and
electromagnetism, which we take to be of the form :
\begin{equation}
{\cal L}_{\phi F} = - \frac{1}{4} B_F(\phi) F_{\mu\nu}F^{\mu\nu}
\end{equation}
where the gauge kinetic function $B_F(\phi)$ is linear, 
\begin{equation}
B_F(\phi) = 1- \zeta \kappa (\phi-\phi_0)
\end{equation}
and $\kappa^2=8\pi G$; $\zeta$ is therefore the relevant coupling, and among other
things it is related to the amount of equivalence principle violations. Constraints on this
coupling are tight at low redshift; conservatively we have \cite{Pospelov,Dvali}:
\begin{equation}
|\zeta_{local}|<10^{-3}\,.\label{localzeta}
\end{equation}
In this work we will derive an analogous but independent constraint coming from the CMB.

This form of the gauge kinetic function can be seen as the first term of a Taylor expansion,
and given the tight low-redshift constraints on varying couplings it is a good approximation
for a slowly varying field at low redshift. Here we are extending it all the way to the CMB epoch.
In this case, how good an approximation this is will be more model-dependent. However, given
that a purely phenomenological parametrization for EDE is already being assumed, this choice
of parametrization for the gauge kinetic function should be quite adequate. We will comment
further on these choices at the end of the paper.

This being said, the evolution of alpha is given by :
\begin{equation}
\frac{\Delta \alpha}{\alpha} \equiv \frac{\alpha-\alpha_0}{\alpha_0} =
\zeta \kappa (\phi-\phi_0) \,,
\end{equation}
and since the evolution of the scalar field can be expressed in terms of the dark energy
properties $\od$ and $w$ as \cite{Couplings1,Couplings2}:
\begin{equation}
w = -1 + \frac{(\kappa\phi')^2}{3 \od} \,,
\end{equation}
(where the prime denotes the derivative with respect to the logarithm of the scale factor)
we finally obtain the following explicit relation specifiying the evolution of the fine
structure constant in the early dark energy class of models :
\begin{equation}
{\alpha / \alpha_0} (a) = 1-\zeta \int_a^{a_0}\sqrt{3\od(a)(1+w(a))}{d\ln{a}}\,.
\end{equation}
As expected the magnitude of the variation is controlled by the strength of the coupling $\zeta$.

\section{Analysis Method}
\label{3}

We have modified the CAMB \cite{camb} code for early dark energy
as in \cite{Calabrese:2010uf} and we include variations in the fine structure 
constant in the recombination process using the method adopted in \cite{avelino} and
modifying the publicly available RECFAST \cite{recfast} routine.

We constrain variation in the fine structure constant $\alpha / \alpha_0$
and $\ome$ by a COSMOMC analysis of the most recent CMB data. The analysis method we adopt is based on the
publicly available Markov Chain Monte Carlo package \texttt{cosmomc}
\cite{Lewis:2002ah} with a convergence diagnostics done through the Gelman and Rubin statistics.

We sample the following eight-dimensional set of cosmological parameters, adopting flat priors
on them: the baryon and cold dark matter densities $\omega_{\rm b}$ and
$\omega_{\rm c}$, the Hubble parameter $H_0$, the scalar spectral index $n_s$,
the overall normalization of the spectrum $A_s$ at $k=0.05$ Mpc$^{-1}$,
the optical depth to reionization, $\tau$, the ratio of the values of the
fine structure constant then and now $\alpha / \alpha_0$ and, finally, the primordial amount of
early dark energy $\ome$. Furthermore, we consider purely adiabatic initial conditions and we impose spatial flatness.

We consider WMAP7 \cite{wmap7} and ACT \cite{act} as CMB datasets and we analyze datasets out to $l_{max}=2500$. In order to investigate the impact of the recent ACT dataset alone we perform two different analysis, WMAP7 and WMAP7+ACT as CMB probes. For the ACT dataset we moreover consider two extra parameters accounting for 
the Poisson and clustering point source foreground components.
We also include informations on dark energy from Baryonic Acoustic Oscillations (BAO) in galaxy surveys \cite{Percival:2009xn},
and finally we impose a prior on the Hubble parameter based on the Hubble Space Telescope observations \cite{hst}. 
More recently, new and improved constraints on the Hubble parameter have been reported in \cite{hst2} and we also perform some analyses considering this new more stringent prior referring to it as HST2.

\section{Cosmology measurements of $\alpha$, $\ome$ and $\zeta$}
\label{4}

\subsection{Constraints from current data}

In Table \ref{tab:wdm} we report the constraints on the $\alpha /\alpha_0$ and 
$\ome$ parameters obtained from the COSMOMC analysis, using the
different combinations of datasets described in the previous section. Notice
the significant improvements on both parameters provided by the small-scale ACT data.
As expected, beacuse of the correlation between $\alpha$ and $H_0$ (see e.g. \cite{menegoni}),
we found an improvement in the constraint on $\alpha$ of about $15 \%$ when HST2 is considered. 
However, when also ACT and BAO are added, the degeneracy is broken and we do not have any relevant contribution from HST2.

\begin{table}[h!]
\begin{center}
\begin{tabular}{l|c|c|c}
\hline
\hline
Datasets & $\alpha / \alpha_0$ & $\ome$ & $\zeta$ \\
\hline
& & & \\
WMAP7+HST & $0.963 \pm 0.044$ & $< 0.064$ &$<0.047$\\
& & & \\
WMAP7+HST2 & $0.960 \pm 0.040$ & $< 0.070$ &$<0.047$\\
& & & \\
WMAP7+ACT+HST & $0.975 \pm 0.020$ & $< 0.060$ &$<0.031$\\
& & & \\
WMAP7+ACT+HST+BAO & $0.986 \pm 0.018$ & $<0.050$ & $<0.025$\\
& & & \\
WMAP7+ACT+HST2+BAO & $0.986 \pm 0.016$ & $<0.050$ & $<0.021$\\
& & & \\
\hline
\hline
\end{tabular}
\caption{Limits at $95 \%$ c.l. on $\alpha / \alpha_0$, $\ome$ and the coupling $\zeta$ from the MCMC anlyses.}
\label{tab:wdm}
\end{center}
\end{table}

\begin{figure}[h!]
\includegraphics[width=8cm]{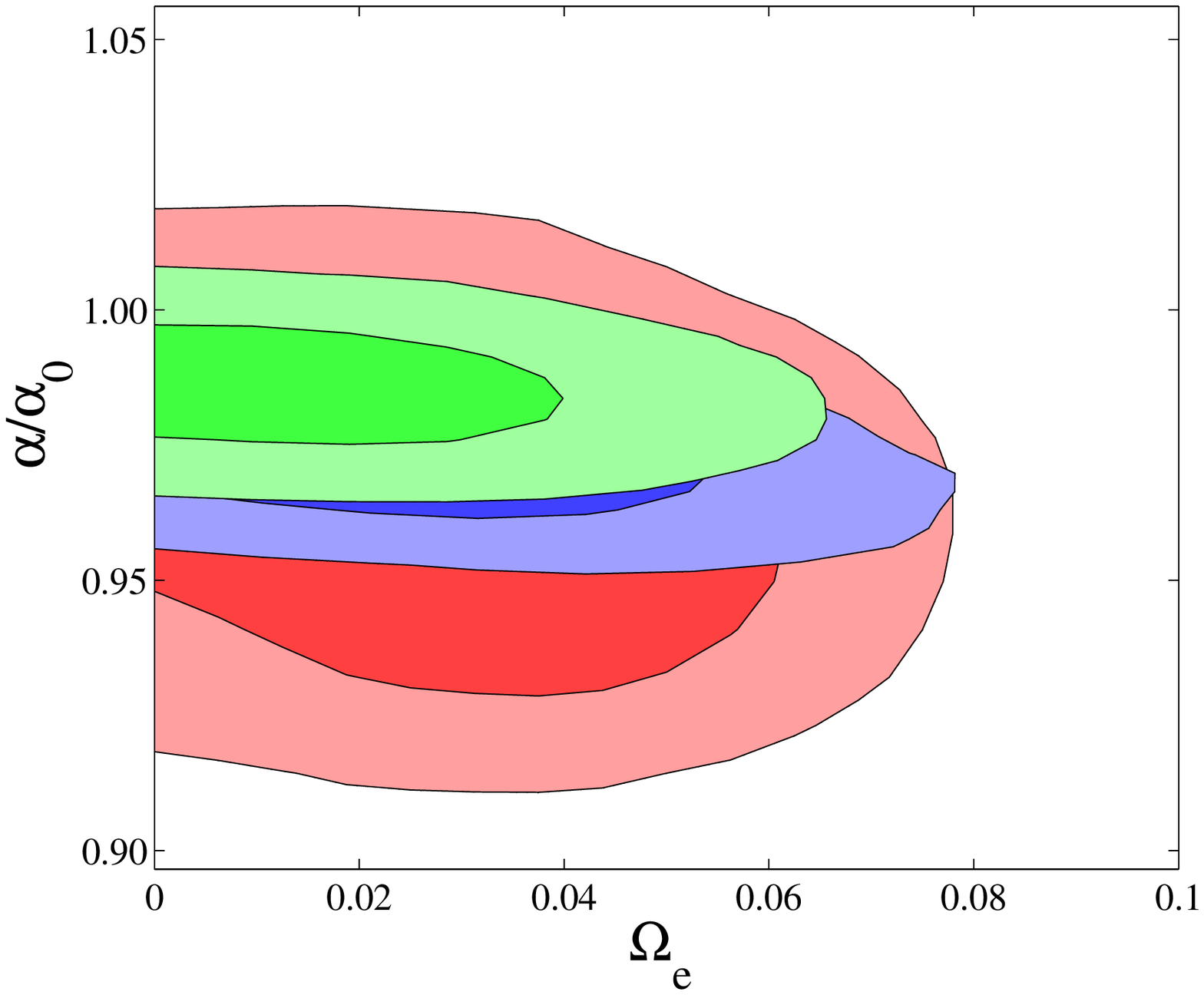}
\caption{\label{plot}$68 \%$ and $95 \%$ c.l. constraints on the $\alpha / \alpha_0$ vs
$\ome$ plane in the case of WMAP7+HST (red contours) compared with WMAP7+ACT+HST (blue contours) and 
WMAP7+ACT+BAO+HST (green contours).}
\end{figure}

In Figure \ref{plot} we show the $68 \%$ and $95 \%$ c.l. constraints on the $\alpha / \alpha_0$ vs $\ome$ plane for different datasets. As we can see, there is no strong degeneracy between $\ome$ and the fine structure constant and the
cosmological data can be used to put strong limits on both quantities.

\begin{figure}[h!]
\includegraphics[width=8cm]{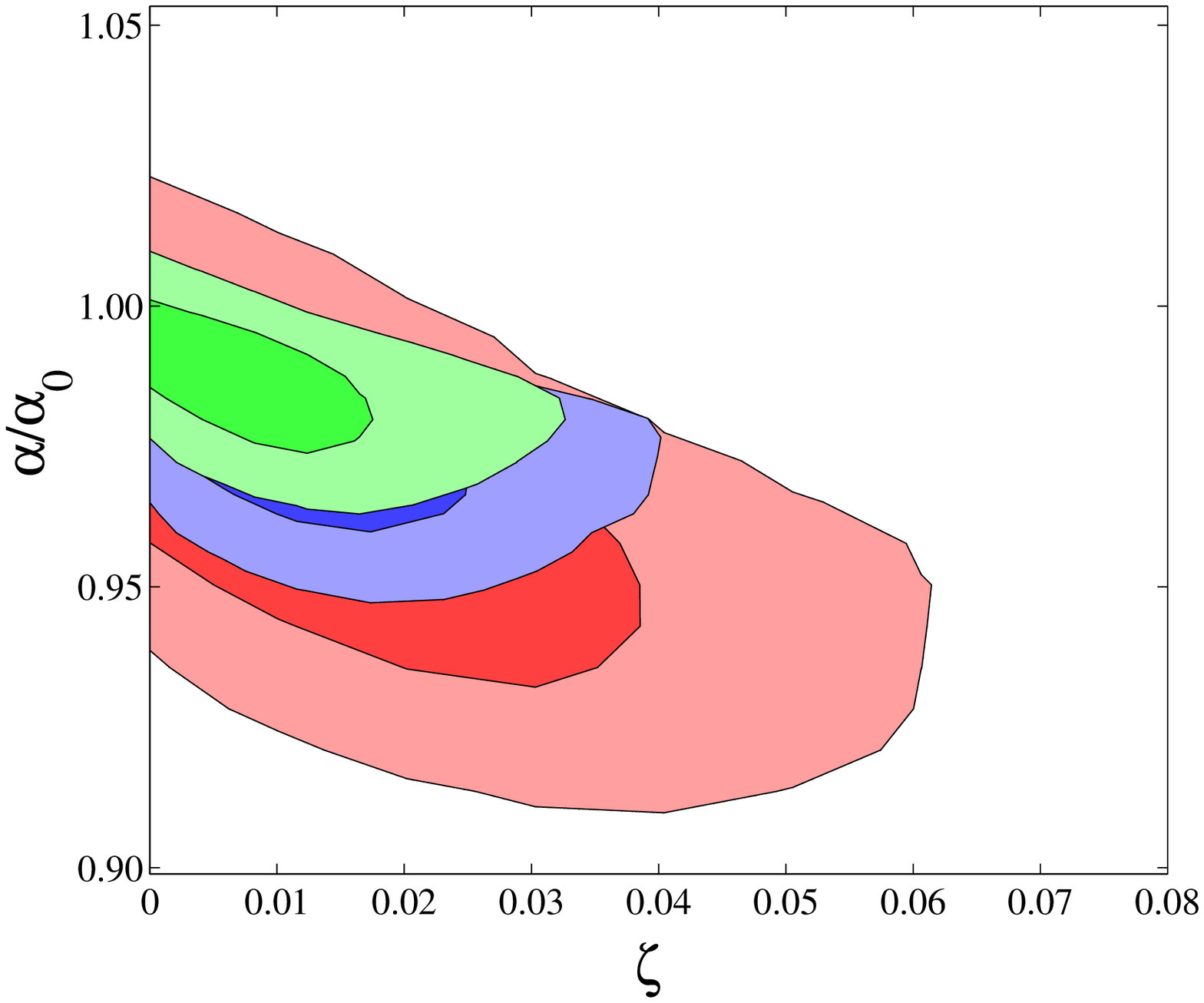}
\caption{\label{plot_coup} $68 \%$ and $95 \%$ c.l. constraints on the $\alpha / \alpha_0$ vs
$\zeta$ plane in the case ofWMAP7+HST (red contours) compared with WMAP7+ACT+HST (blue contours) and 
WMAP7+ACT+BAO+HST (green contours).}
\end{figure}

We can therefore consider the value of the coupling parameter at decoupling:

\begin{equation}
\zeta= (1-{\alpha / \alpha_0})/ \int_{a_{dec}}^{a_0}\sqrt{3\od(a)(1+w(a))}{d\ln{a}}
\end{equation}
and use the constraints on $\ome$ and ${\alpha / \alpha_0}$ obtained in our analysis
to constrain variations in $\zeta$. In Table I we report these constraints for
the different analysis while Figure \ref{plot_coup} shows the degeneracy 
between $\alpha / \alpha_0$ and $\zeta$.

Although the current constraints are 20-40 times weaker than the ones that can be obtained from weak equivalence principle tests (cf. Eqn. \ref{localzeta}), it's important to  keep in mind that our constraints are obtained on completely different scales, cosmological ones, as opposed to laboratory ones. So a discrepancy of less than two orders of magnitude is actually impressive.

To put this in context, in the case of the Eddington parameter $\gamma$, the difference in sensitivity between cosmological measurements (obtained from lensing \cite{Schwab}, thus on kiloparsec scales) and solar system ones (from the Cassini bound \cite{Bertotti}, effectively on $10^{-4}$ parsec scales) is more than three orders of magnitude. This difference testifies to the exquisite sensitivity of current CMB datasets and of course the situation will improve with forthcoming datasets, as we now discuss.

\subsection{Forecasts for Planck and CMBPol}

To evaluate future sensitivity to these parameters from CMB we consider noise properties consistent
with the Planck~\cite{planck} and CMBPol \cite{Bock:2009xw} experiments (see Table~\ref{tab:exp} for specifications).

\begin{table}[htb!]
\begin{center}
\begin{tabular}{lcccc}
& & & & \\
\hline
\hline
Experiment \ & Channel \ & FWHM & $\ \ \Delta T \ \ $ & $\Delta P$\\
\hline
Planck & 70 & 14' & 12.8 & 18.3\\
\phantom{Planck} & 100 & 10' & 6.8 & 10.9\\
\phantom{Planck} & 143 & 7.1'& 6.0 & 11.4\\
\hline
CMBPol & 70 & 12.0' & 0.148 & 0.209\\
\phantom{Planck} & 100 & 8.4' & 0.151 & 0.214\\
\phantom{Planck} & 150 & 5.6' & 0.177 & 0.250\\
\hline
$f_{sky}=0.85$ & & & \\
\hline
\hline
\end{tabular}
\caption{Planck and CMBPol experimental specifications. Channel frequency is given in GHz, Full-Width at Half-Maximum (FWHM) in arc-minutes, and the temperature and polarization sensitivity per pixel in $\mu K$.}
\label{tab:exp}
\end{center}
\end{table}

We consider for each channel a detector noise of $w^{-1} = (\theta\sigma)^2$, where $\theta$ is the Full-Width at Half-Maximum
(FWHM) of the beam assuming a Gaussian profile and $\sigma$ is the temperature/polarization sensitivity $\Delta T$/$\Delta P$ (see Tab.~\ref{tab:exp}). We therefore derive a noise spectrum given by:
\begin{equation}
N_\ell = w^{-1}\exp(\ell(\ell+1)/\ell_b^2) \, ,
\end{equation}
where $\ell_b$ is $\ell_b \equiv \sqrt{8\ln2}/\theta$.

We then perform a Fisher matrix analysis \cite{Bond:1997wr} to estimate the $1-\sigma$ error for each parameter. 
We assume a $\Lambda$CDM fiducial model with the following parameters: $\Omega_b h^2=0.02258$, $\Omega_c h^2=0.1109$, $\tau=0.088$, $H_0=71\, {\rm km/s/Mpc}$, $n_s=0.963$ plus the EDE parameters that we fix to : $w_0=-0.90$, $\ome=0.03$, $c_{s}^2=1$, $c_{vis}^2 = 0$ and $\alpha/\alpha_0 =1$.

Results for the parameters of direct interest are shown in Table \ref{fish}. Forecasts for the fine-strcture constant are comparable to those of our earlier work \cite{Rocha}, when one allows for the slightly different assumptions and fiducial model. More interestingly, we see that Planck can constrain the scalar field coupling $\zeta$ with an accuracy comparable to that of current local bounds, while CMBPol can do about five times better. However, it should also be ponted out that local bounds are expected to improve with the advent of satellite-based tests of the equivalence principle.

\begin{table}[h!]
\begin{center}
\begin{tabular}{l|c|c|c}
\hline
\hline
Experiment & $\sigma_{\alpha / \alpha_0}$ & $\sigma_{\ome}$ & $\sigma_{\zeta}$ \\
\hline
& & & \\
Planck & $0.0012$ & $0.0036$ &$< 0.0012$\\
& & & \\
CMBPol & $0.00025$ & $0.0015$ & $< 0.00022$\\
& & & \\
\hline
\hline
\end{tabular}
\caption{Fisher matrix errors at $68 \%$ c.l. on $\alpha / \alpha_0$ and $\ome$ and upper bounds at $95 \%$ on coupling $\zeta$ from Planck and CMBPol.}
\label{fish}
\end{center}
\end{table}

\begin{figure}[h!]
\includegraphics[width=7.2cm]{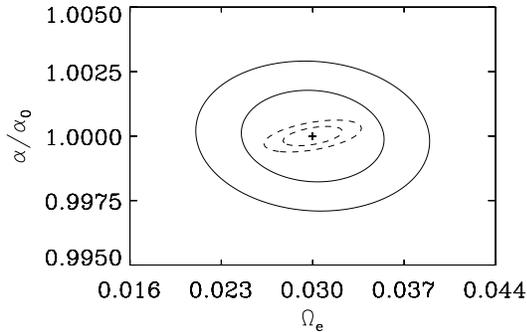}
\caption{\label{fisher} $68 \%$ and $95 \%$ c.l. estimated constraints on the $\alpha / \alpha_0$ vs
$\ome$ plane from Planck (solid lines) and CMBPol (dashed lines).}
\end{figure}

We report in Figure \ref{fisher} the 2-dimensional contour plots in the $\alpha/\alpha_0 - \ome$ plane showing the $1$ and $2 - \sigma$ regions for Planck with solid lines and CMBPol with dashed lines. The effect of the better polarization data in further constraining $\alpha$ is quite noticeable.

\section{Conclusions}

We have discussed present constraints on variations of the fine structure constant $\alpha$ 
in the context of phenomenological class of early dark energy where the putative dynamical scalar field is coupled to the electromagnetic sector of the theory through the simplest allowed (linear) coupling. We have found no significant correlation between $\alpha / \alpha_0$ and
the early dark energy density parameter $\Omega_e$, providing the
constraints $\alpha / \alpha_0 = 0.975 \pm 0.020$ and $\Omega_e < 0.060$ at $95 \%$ c.l.
The constraints on $\alpha$ are significantly improved with respect to the
WMAP five-years constraint \cite{menegoni}, thanks to the new ACT data.

Moreover we have used the CMB and other cosmological datasets to also constrain the dimensionless parameter quantifying the strength of the scalar field coupling, $\zeta$, at the epoch of decoupling, in addition to constraining $\alpha$ itself. While the current data only allowed us to obtain a constraint on $\zeta$ that is about 20 times weaker than those obtainable locally (which come from Equivalence Principle tests), we emphasize that the two contraints are independent. Moreover, forthcoming or future missions, such as Planck Surveyor and CMBPol, can match and possibly even surpass the sensitivity of current local tests.

Admittedly the toy model which we have studied may well be too simplistic, and we are currently investigating how sensitive our results are to our choice of parametrization. Nevertheless, we emphasize that the model has the clear virtue of being predictive enough to be confirmed or ruled out by forthcoming experiments. This will require a combination of cosmological datasets (Planck being the best example) and better low-redshift constraints on variations of $\alpha$, which will soon be available.

Finally, this work is an example of how one can use the early universe as a laboratory in which to directly probe fundamental physics scenarios, in ways that will complement and enhance local, small-scale tests. Moreover, further avenues for searching for new physics will soon be available. Now that experiments such as MICROSCOPE\cite{miscroscope} and ACES \cite{aces} are getting ready to perform Equivalence Principle tests in microgravity, it is encouraging that there are good prospects for carrying out analogous tests on astrophysical and cosmological scales. Synergies between local experiments and cosmological observations will soon allow us to probe unexplored aspects of gravity and fundamental physics. We will revisit these issues in future work.

\begin{acknowledgments}
This work is supported by PRIN-INAF, "Astronomy probes fundamental physics".
The work of C.M. is funded by a Ci\^encia2007 Research Contract, funded by FCT/MCTES (Portugal) and POPH/FSE (EC).
C.M. and G.R. also acknowledge additional support from project PTDC/FIS/111725/2009 from FCT, Portugal.
We thank Eric V. Linder for useful comments.
\end{acknowledgments}

\end{document}